\documentstyle[12pt]{article}
\begin{document}
E-print: hep-th/0008129
\vskip1cm
\centerline{\large Supersymmetric Superconducting Bag as  }
\centerline{\large a Core of Kerr Spinning Particle}
\vskip1cm
\par
\centerline{A. Ya. Burinskii}
\smallskip
\centerline{NSI of Russian Academy of Sciences }
\centerline{B. Tulskaya 52  Moscow 113191 Russia,
 e-mail:bur@ibrae.ac.ru}
\vskip1cm
\begin{quotation}
The problem of a regular matter source for the Kerr spinning particle
is discussed. A class of minimal deformations of the Kerr-Newman solution
is considered obeying the conditions of regularity and smoothness for the
metric and its matter source.
\par
It is shown that for charged source corresponding matter forms a rotating
bag-like core having (A)dS interior and smooth domain wall boundary.
Similarly, the requirement of regularity of the Kerr-Newman electromagnetic
field leads to superconducting properties of the core.
\par
    We further consider the U(I) x U'(I) field model ( which was used
by Witten to describe cosmic superconducting strings ), and we show that it
can be adapted for description of superconducting bags having a long range
external electromagnetic field and another gauge field confined
inside the bag. Supersymmetric version of the Witten field model
given by Morris is analyzed, and corresponding BPS domain wall
solution interpolating between the outer and internal supersymmetric
vacua is considered. The charged bag bounded by this BPS domain wall
represents an `ultra-extreme' state with a total mass which is lower
than BPS bound of the wall. It is also shown that supergravity suggests
the AdS vacuum state inside the bag.
\par
     Peculiarities of this model for the rotating bag-like source of the
Kerr-Newman geometry are discussed.
\end{quotation}
\newpage
\section{Introduction}
The Kerr solution is well known as  a  field  of  the
rotating black hole. However, for the case of  a  large  angular
momentum $l$; $\mid a\mid  = l/m \geq  m$  all the horizons of the Kerr
metric are absent, and there appears a naked  ring-like  singularity.
This singularity has many unpleasant manifestations and must be regularized
by being hidden inside a matter source.
The Kerr solution  with $\mid
a\mid  \gg  m$   displays some remarkable features  indicating  a relation
to the structure  of   the   spinning   elementary particles.
\par
In 1969 Carter \cite{Car} observed,  that  if three parameters of the
Kerr-Newman solution are adopted to be ($\hbar $=c=1 )
\begin{equation}
\quad e^{2}\approx  1/137,\quad
m \approx 10^{-22},\quad a \approx  10^{22},\quad ma=1/2,
\label{par}
\end{equation}
then one obtains  a  model  for  the four  parameters  of  the electron:
charge $e$, mass $m$, spin $l$ and  magnetic moment $ea$,
and the gyromagnetic ratio is automatically the same as that of the Dirac
electron.  The first treatment of the disc-like source of the Kerr spinning
particle was given by Israel \cite{Is}
in the form of an infinitely thin disc spanned by the Kerr singular
ring \footnote{Disc has the Compton size with radius $a=l/m=\frac 1{2m}$}.
   Then, some stringy structures were obtained in Kerr geometry  \cite{str},
and the Israel results where corrected by Hamity \cite{ham}
showing that the disc is to be in a rigid relativistic rotation. Further,
L\`opez \cite{lop} suggested a regularized model of the source
constructing the disc in the form of a (infinitely thin) rotating elliptical
shell ( bubble ) covering the singular ring.
Besides the bubble models, the solid disc-like
sources were considered, too \cite{Kr,solid}.
The structure of the electromagnetic field near the disc suggested
superconducting properties of the material of the source
\cite{tiom,lop0,solid},
and there was obtained an analogue of the Kerr singular ring with
the Nielsen-Olesen \cite{NO} and Witten \cite{Wit} superconducting strings.
Since 1992 there has been considerable interest as to black holes in string
theory and the point of view that some of black holes can be
treated as elementary particles \cite{part,Sen,BS,c-str}.
  In particular, Sen \cite{Sen} has obtained a generalization of the Kerr
solution to low energy string theory, and it was shown \cite{BS} that near
the Kerr singular ring the Kerr-Sen solution acquires a metric similar to
the field around a heterotic string.
\footnote{One should also mention a super-version of the Kerr-Newman
solution \cite{super-K,superBH} involving fermionic fields which are
important for a complete description of spinning particle.}
Therefore, it was obtained that the structure of the Kerr source contains
some elements of the known extended sources such as strings, membranes,
bubbles and domain walls.
However, alongside with singular description of such extended sources,
there exists a regular soliton-like description based on the nonlinear field
models with broken symmetry, and in particular, on the Higgs model of
superconductivity.
\par
 The aim of this paper is to demonstrate that a  smooth bag-like
source  appears in the core of Kerr geometry as a consequence
of simple and natural proposals on the regularity of metric bounded by the
Kerr-Schild class, and to represent a field model suitable for the
description of the properties of the particle-like source of the Kerr-Newman
field.
Analyzing the known reasonable smooth sources based on the Higgs models,
we show that only the $U(I)\times \tilde U(I)$ field
model (used by Witten for superconducting cosmic strings \cite{VS, Wit})
possesses the external long range electromagnetic field which is necessary
to describe the Kerr-Newman source.
We further consider the supersymmetric version of this field model given by
Morris \cite{Mor} and show that it can be adapted for description of the
bag-like superconducting core of the Kerr-Newman field. The supersymmetric
vacuum states of this model are analyzed, and an approximate bag-like domain
wall solution separating the internal ( false ) and external ( true )
vacua is considered. We show that this domain wall represents a BPS
saturated state. It is remarkable that the total mass of the charged bag
formed from this BPS saturated domain wall turns out to be lower than BPS
energy bound for this domain wall ( so called `ultra-extreme' state
\cite{CvSol} ).
Therefore, the way to overcome the BPS bound \cite{GH} is opened in this
model, which is necessary to get the real ratio of parameters (\ref{par}).
\par
	We should note that similar problem of the metric regularization
appears in another context as the problem of a regular black hole interior
\cite{FMM,Dym,PI,Mag,EH,BEHM}. Many of the results obtained in this direction
as well as the methods of treatment resemble those of the
problem of the Kerr spinning particle.
In particular, in the case of the absence of the horizons the core is
treated as a visible semi-closed world, there also appear bubbles with
infinitely thin domain walls \cite{FMM,PI}, and a crucial
role of the demand of regularity for the Kerr-Schild class of metrics was
mentioned \cite{EH}. In some treatments, for example in \cite{Dym},
the relation of both of the problems is also discussed. However, when
considering the matter
field source we apply not to the corrections of stress-energy tensor
caused by quantum fields of vacuum polarization but rather to the
supersymmetric vacuum states caused by a multiplet of Higgs fields in
supergravity. As a result, we come to the conclusion that the case of
AdS internal space inside the charged core of the Kerr-Newman geometry
looks the most plausible.
\par
\section{Stress-energy tensor of the Kerr source}
\par
The Kerr-Newman metric in the Kerr-Schild form is
\begin{equation}
g_{\mu\nu} =\eta_{\mu\nu} +2 h k _\mu k _\nu,
\label{g}
\end{equation}
where $\eta_{\mu\nu}= diag (-1,1,1,1)$ is an auxiliary Minkowski space,
$ h =\frac {f(r)}{r^2+a^2 \cos ^2 \theta}$,  $k_\mu$ is a vector
field tangent
to the Kerr geodesic and shear free congruence \cite{DKS},~\footnote{Its
concrete form will not be
essential for our consideration in this paper.} and function $f$ has the
Kerr-Newman form $f_{KN}(r)=mr-e^2/2$, where $r$ and $\theta$ are oblate
spheroidal coordinates ( family of confocal ellipsoids $ r=const.$ covers
the disc $r=0$ ). This metric corresponds to a regular electrovacuum
space-time for the exclusion of singular ring at $r=\cos\theta=0$.
This ring is a branch line of space on two sheets corresponding to
$r>0$ and $r<0$.
\par
Previous attempts to introduce the Kerr source were connected
with a truncation of the `second' sheet along the disc $r=0$. This procedure
breaks the analyticity of the space, and there appears a singular
distribution
of material sources on the disc, as a consequence of the Einstein
field equations. The choice of truncation is not unique, and
diverse singular models of the Kerr source were considered
\cite{Is,lop,Kr}.
\par
Another approach to this problem is connected with a smooth deformation of
the Kerr-Newman space in a neighborhood of the Kerr disc  \cite{lop1,GG,BEHM}
keeping the Kerr-Schild form of metric.
The deformed metric induces a stress-energy tensor in the right side of the
Einstein equations which corresponds to the appearance of some extra matter
surrounding the disc $r=0$, and there appears an image of the source of
the Kerr-Newman geometry in the form of a rotating disc with a smooth
matter distribution.
\par
On the basis of the results of recent calculations \cite{BEHM} performed in the
Kerr-Schild formalism \cite{DKS}, we would like to show here that imposing
some minimal
conditions of regularity and smoothness on the resulting metric and energy
distribution one can obtain a shape of the source in the form of a rotating
bag with a smooth domain wall boundary.
\par
Following \cite{BEHM}, we assume that the deformed metric retains the
Kerr-Schild form (\ref{g}) and the form of the Kerr principal null congruence
$k_{\mu} (x)$ while the function  $f(r)$ is restricted by the following
conditions:
\par
  i)  the deformed metric has to be regular and smooth for $r\ge 0$ and
      has to lead to a regular matter distribution in core region;
\par
  ii)  when out of the core, function $f(r)$ has to take the
  Kerr-Newman form $f_{KN}(r)=mr-e^2/2$, where $m$ and $e$ are the total
  mass and charge parameters.
\par
To satisfy the condition i) for small values of $r$,  leading term of
the function $f(r)$ has to be at least of the order $\sim r^n$ with $n\ge 4$.
The analysis of the stress-energy tensor for  metrics
(\ref{g}), given in \cite{BEHM},
selects some peculiar properties of the case $n=4$.
In this case  $f(r)=f_0(r)=\alpha r^4$ in the core region, and
in the nonrotational case ($ a=0 $ ) space-time has a constant curvature
generated by a homogenous matter distribution with energy  density
$\rho= \frac 1{8\pi} 6\alpha$ in the core.
This fact motivates one more condition:
\par
 iii) the internal region of the core has to be described by function
$f_0(r) =\alpha r^4$.
\par
As a result of ii) and iii) the boundary of core $r_0$ may be estimated
as a point of intersection of $f_0(r)$ and  $f_{KN}(r)$
and  the resulting smooth function $f(r)$ must be interpolating between
functions $f_0(r)$ and $f_{KN}(r)$ near the boundary of the core
$r\approx r_0$.
\par
It is surprising that the bag-like
structure of the Kerr-Newman source is a direct result of these simple and
 natural conditions.
\par
The analysis given in \cite{BEHM} shows that metric
(\ref{g}) can be expressed via orthonormal tetrad as follows
\begin{equation}
g_{\mu\nu} =m_\mu m_\nu + n_\mu n_\nu + l _\mu l_\nu - u_\mu u_\nu,
\label{gBL}
\end{equation}
and the corresponding stress-energy tensor of the source following from the
Einstein equations may be represented in the form
\begin{equation}
T_{\mu\nu}^{(af)} = (8\pi)^{-1} [(D+2G) g_{\mu\nu} -
(D+4G) (l_\mu l_\nu -  u_\mu u_\nu)],
\label{Taf}
\end{equation}
where
\begin{equation}
u_\mu= - \sqrt {\frac{\Delta}{\Sigma}} (1,0,0,-a\sin^2 \theta)
\label{u}
\end{equation}
is the unit time-like four-vector,
\begin{equation}
l_\mu =\sqrt{\frac{\Sigma}{\Delta}}(0,1,0,0)
\label{l}
\end{equation}
is the unit vector in radial direction, and
\begin{equation}
n_\mu =\sqrt{\Sigma}(0,0,1,0),
\label{n}
\end{equation}
\begin{equation}
m_\mu =\frac{\sin\theta}{\sqrt{\Sigma}}(a,0,0, -r^2 -a^2)
\label{m}
\end{equation}
are two more space-like vectors.
Here $\Delta= r^2 +a^2 -2f$ and $\Sigma = r^2 + a^2 \cos ^2 \theta$,
\begin{equation}
 D= - f^{\prime\prime}/(r^2 + a^2 \cos ^2 \theta),
\label{D}
\end{equation}
\begin{equation}
G= (f'r-f)/ (r^2 + a^2 \cos ^2 \theta)^2,
\label{G}
\end{equation}
and
the Boyer-Lindquist coordinates $t,r,\theta, \phi$ are used.
The expressions (\ref{Taf}), (\ref{D}), (\ref{G})
show that the source represents a smooth
distribution of rotating confocal ellipsoidal layers $r=const.$, and
stress-energy tensor has the form corresponding to an anisotropic fluid.
\footnote{Similar interpretation was also given in \cite{lop1}, however
the expression for stress-energy tensor is not correct there, it does not
contain $D$ term. The form (\ref{Taf}) is close to the expression obtained
in \cite{GG}.}
\par
Like to the results for singular (infinitely thin) shell-like source
\cite{ham,lop}, the stress-energy tensor can be diagonalized in a comoving
coordinate system showing that the source represents a relativistic rotating
disc. However, in this case, the disc is separated into
ellipsoidal layers each of which rotates rigidly with its own angular
velocity $\omega (r) =a/(a^2+r^2)$.
In the comoving coordinate system the tensor $T_{\mu\nu}$ takes the form
\begin{equation}
T_{\mu\nu} =  \frac{1}{8\pi}
\left( \begin{array}{cccc}
2G&0&0&0 \\
0&-2G&0&0 \\
0&0&2G+D&0 \\
0&0&0&2G+D
\end{array} \right),
\label{Tful}
\end{equation}
that corresponds to energy density
$\rho = \frac{1}{8\pi} 2G$,
 radial pressure $p_{rad} = -\frac{1}{8\pi} 2G$,
and tangential pressure $p_{tan} = \frac{1}{8\pi}(D+ 2G)$.
\par
Setting $a=0$ for the non-rotating case, we obtain
$\Sigma= r^2$, the surfaces $r=const.$ are spheres and we have spherical
symmetry for all the above relations. The region described by
$f(r)=f_0(r)$ is the region of constant value of the scalar curvature
invariant $R=2D= - 2 f_0^{\prime\prime}/r^2 = - 24\alpha$, and of constant
value of energy density
\begin{equation}
T_{\mu\nu} =  \frac{1}{8\pi} 6\alpha
\left( \begin{array}{cccc}
1&0&0&0 \\
0&-1&0&0 \\
0&0&-1&0 \\
0&0&0&-1
\end{array} \right).
\label{Tbal}
\end{equation}
\par
If we assume that the region of a constant curvature is closely extended
to the boundary of source $r_0$ which is determined as a root of
the equation
\begin{equation}
f_0(r_0)=f_{KN}(r_0),
\label{r0}
\end{equation}
then, smoothness of the $f(r)$ in a
small neighborhood of $r_0$, say $\vert r-r_0 \vert <\delta$, implies
a smooth interpolation  for the derivative of the function $f(r)$
between $f^\prime _0(r)\vert _{r=r_0-\delta}$ and
$f^\prime _{KN}(r)\vert_{r=r_0+\delta}$.
Such a smooth interpolation on a small distance $\delta$ shall
lead to a shock-like increase of the second derivative
$f^{\prime\prime} (r)$ by $ r \approx r_0 $.
Graphical analysis shows that the result will be different for charged and
uncharged cases.
\par
In charged case for $\alpha \le 0$ there exists only one positive root $r_0$,
and second derivative of the smooth function $f^{\prime\prime} (r)$
is positive near this point. Therefore, there appears an extra tangential
stress near $r_0$ caused by the term
$D=- f^{\prime\prime} (r)/(r^2 + a^2 \cos ^2 \theta )\vert _{r=r_0}$
in the expression (\ref{Tful}).
It can be interpreted as the appearance of an effective shell ( or a domain
wall ) confining the charged ball-like source with a geometry of a constant
curvature inside the ball.
The case
$\alpha=0$ represents the bubble with a flat interior which has
in the limit $\delta\to 0$ an infinitely thin shell. It
corresponds to the Dirac electron model in the case of a spherical shell and
to the L\`opez rotating shell model for a spinning particle \cite{lop,lop2}.
\par
The point $r_e=\frac {e^2}{2m}$ corresponding to a ``classical size" of
electron is a peculiar point as a root of the equation $f_{KN}(r) =0$.
 It should be noted that by $\alpha=0$ the equation (\ref{r0}) yields
the root $r_0=r_e$. For $\alpha <0$ position of the roots is $r_0<r_e$,
and for $\alpha >0$ one obtains $r_0> r_e $. One can also see that
 $r_0 \to 0$ by $\alpha\to -\infty$.
Therefore, there appear four parameters characterizing the function
$f(r)$ and the corresponding bag-like core: parameter $\alpha$ characterizing
cosmological constant inside the bag $\Lambda _{in}= 6 \alpha$,
two peculiar points $r_0$ and $r_e$ characterizing the size of the bag and
parameter $\delta$ characterizing the smoothness of the function $f(r)$ or
the thickness of the domain wall at the boundary of core.
\par
For $\alpha > 0$ two real roots of (\ref{r0}) can exist.
If $r_0^{(1)} < r_0^{(2)}$ then only near the root $r_0^{(1)}$ the
second derivative $f^{\prime\prime}(r)$ is positive, and the confining shell
appears.
However, if the root $r_0^{(2)}$ is chosen as a boundary of the source
then the second derivative $f^{\prime\prime} (r)$ is negative by
$ r \approx r_0 $, and a tangential pressure appears
on the boundary of source instead of the stress.
\par
The internal geometry of the ball is de Sitter one for $\alpha>0$,
anti de Sitter one for $\alpha <0$ and flat one for $\alpha=0$.
One should note that this effective shell has a more sharp manifestation
for AdS interior ($\alpha<0$), and it can be absent at all for charged
source by $\alpha>0$.
\par
For the uncharged source Eq.(\ref{r0}) has the root $r_0=0$,
and by $\alpha \le 0 $ there are no positive roots at all. If $\alpha >0$
there is one positive root $r_0$ and the second derivative
$f^{\prime\prime} (r)$ is negative there leading to the appearance of an
extra tangential pressure on the boundary of source instead of the stress.
Therefore, the confining shell is absent, and the source is
confined by the radial stress coming from the term $G$ that is positive
in this case.
\par
Let us consider peculiarities of the rotating Kerr source.
In this case $\Sigma= r^2 + a^2 \cos ^2 \theta$, and the
surfaces $r=const.$ are ellipsoids described by the equation
\begin{equation}
\frac{x^2 +y^2}{r^2 + a^2} + \frac {z^2}{r^2} =1.
\label{ell}
\end{equation}
Energy density inside the core will be constant only in the equatorial plane
$\cos \theta=0$. Therefore, the Kerr singularity is regularized and the
curvature is constant in string-like region $r<r_0$ and $\theta=\pi/2$
near the former Kerr singular ring. At the same time, when considering
the energy density in central regions of
the disc corresponding to $\theta=0$,
 and taking into account
the values of parameters for spinning particle (\ref{par}),
one finds that its ratio to the energy density at $\theta=\pi/2$ will be
negligible
$\frac{\rho \vert_{\theta=0}}{\rho\vert _{\theta=\pi/2}}\approx
r^2/ a^4 < r_e^2/a^4 \sim 10^{-48}$.
Similarly, estimating the ratio $2G/D$ near the axis of the disc one finds
that energy density is also negligible in respect to the stress of
domain wall in this region
$\vert 2G/D\vert \sim \vert 2(f^{\prime} r-f) /f^{\prime\prime}a^2 \vert <
(r_e/a)^2 <10^{-4}$
and the ratio
$\frac{stress\vert _{\theta=0}}{stress \vert_{\theta=\pi/2}}
< (r_e/ a)^4 = e^8 < 10^{-8}$ shows a
strong increase of the stress near the string-like boundary of the disc.
\par
\section{Field model for the bag-like Kerr source}
\par
The known models of the bags \cite{bag} and cosmic bubbles
\cite{Mac,MB} with smooth domain wall boundaries are based on the Higgs
scalar field $\phi$  with a Lagrange density of the form
$L=- \frac 12 \partial _\mu \phi\partial ^\mu \phi
-\frac {\lambda ^2}8 (\phi ^2-\eta^2)^2 $ leading to the kink
planar solution ( the wall is placed in $xy$-plane at $z=0$ )
\begin{equation}
\phi (z)= \eta \tanh (z/\delta),
\label{kink}
\end{equation}
where $\delta =\frac 2{\lambda \eta}$ is the wall thickness.
The kink solution describes two topologically distinct vacua
$<\phi>= \pm \eta$ separated by the domain wall.
\par
The stress--energy
tensor of the domain wall is
\begin{equation}
T_{\mu}^{\nu} = \frac{\lambda ^2 \eta ^4}4 \cosh ^{-4}(z/\delta)
diag (1,1,1,0),
\label{Tkink}
\end{equation}
indicating a surface stress within the plane of the wall which is equal
to the energy density.
When applied to the spherical bags or cosmic bubbles \cite{bag,Mac,MB},
the thin wall
approximation is usually assumed $\delta \ll r_0$, and a spherical domain
wall separates a false vacuum inside the ball ($r<r_0$) $<\phi>_{in}=-\eta$
from a true outer vacuum $<\phi>_{out}=\eta$.
\par
In the gauge string models \cite{NO}, the Abelian Higgs field provides
confinement of the magnetic vortex lines in superconductor. Similarly,
in the models of superconducting bags \cite{SkS, dual},
the gauge Yang-Mills or quark fields are confined
in a bubble ( or cavity ) in superconducting QCD-vacuum.
\par
A direct application of the Higgs
model for modelling superconducting properties of the Kerr source
is impossible since the Kerr source has to contain the external long range
Kerr-Newman electromagnetic field, while in the models of strings and bags
the situation is quite opposite: vacuum is superconducting in external
region and electromagnetic field acquires a mass there from Higgs field
turning into a short range field.
An exclusion represents the  $U(I)\times \tilde U(I)$ cosmic string model
given by Vilenkin-Shellard and Witten \cite{VS,Wit}  which
represents a doubling of the usual Abelian  Higgs model.
The model  contains
two sectors, say $A$ and $B$, with two Higgs fields $\phi_A$ and $\phi_B$,
and two gauge fields $A_{\mu}$ and $B_{\mu}$ yielding two sorts
of superconductivity $A$ and $B$.
It can be adapted to the bag-like source in such a manner
that the gauge field $A_{\mu}$ of the $A$ sector has to describe a
long-range electromagnetic field in outer region of the bag
while the chiral scalar field of this sector $\phi_A$ has to form a
superconducting core inside the bag which must be unpenetrable
for $A_{\mu}$ field.
\par
The sector $B$ of the model has to describe the opposite situation.
The chiral field $\phi_B$ must lead to a $B$-superconductivity in outer
region confining the gauge field $B_{\mu}$ inside the bag.
\par
The corresponding Lagrangian of the Witten $U(I)\times \tilde U(I)$
field model is given by
\footnote{We use signature $(-+++)$.}
\cite{Wit}
\begin{equation}
L=-(D^\mu \phi_A )(\overline{ D_\mu \phi_A} )-(\tilde D^\mu \phi_B )
(\overline{\tilde D_\mu \phi_B} )-\frac 14 F_A^{\mu \nu }F_{A\mu \nu }-
\frac 14 F_B^{\mu \nu } F_{B\mu \nu}-V,
\label{L}
\end{equation}
where
$F_{A\mu \nu }=\partial _\mu A_\nu -\partial _\nu A_\mu $ and
$F_{B\mu \nu}=\partial _\mu B_\nu -\partial _\nu B_\mu $
are field stress tensors, and the potential has the form
\begin{equation}
V=\lambda (\bar{\phi_B}\phi_B -\eta ^2)^2+f(\bar{\phi_B}
\phi_B -\eta ^2)\bar{\phi_A}\phi_A +m^2\bar{\phi_A}\phi_A +\mu
(\bar{\phi_A}\phi_A )^2.  \label{V}
\end{equation}
Two Abelian gauge fields $A_\mu$ and $B_\mu$ interact separately
with two complex scalar fields $\phi _B$ and $\phi_A$ so that the
covariant derivative $D_\mu \phi_A =(\partial +ie A_\mu) \phi_A $ is
associated with $A$ sector, and covariant derivative
$\tilde D_\mu \phi _B = ( \partial +ig B_\mu) \phi _B $ is associated with
$B$ sector.
Field $\phi _B $ carries a $\tilde U(1)$ charge $\tilde q \ne 0$ and a
$U(1)$ charge $q=0$,
 and field $\phi_A $ carries $\tilde U(1)$ and $U(1)$ charges of
$\tilde q=0$ and $q\ne 0$, respectively.
\par
Outside the string core, the group $U(I)$ is unbroken and $A_\mu$
describes a long range electromagnetic field. At the same time
group $\tilde U(I)$ is broken outside the string.
Within the string core group $U(I)$ gets broken giving rise to
superconducting condensate and nonvanishing current.
Parameters $\lambda $, $\mu $, $f$, and $m$ are to be positive.
The vacuum state outside the core is characterized by
$\bar \phi_B \phi_B =\eta ^2$ and $\phi_A =0$.
In the string core $\phi_B= 0$, and
in the range of parameters $(f\eta^2-m^2)>0$, there appears a
superconducting $\phi_A$-condensate $\vert\phi_A \vert = \phi_0 >0$
breaking the gauge group $U(I)$ and generating the mass $m_A=e^2 \phi_0 ^2$
for the electromagnetic field.
\par
Therefore, the model fully retains the properties of the usual bag models
which are described by $B$ sector providing confinement of $B_{\mu}$ gauge
field inside bag, and it acquires the long range electromagnetic field
$A_{\mu}$ in the outer-to-the-bag region described by sector $A$.
The A and B sectors are almost independent interacting only
through the potential term for scalar fields. This interaction has to
provide synchronized phase transitions from superconducting B-phase
inside the bag to superconducting A-phase in the outer region.
The synchronization of this transition occurs explicitly in a supersymmetric
version of this model given by Morris \cite{Mor}.
\par
\section{Supersymmetric Morris model}
\par
In Morris model, the main part of Lagrangian of the bosonic sector is
similar to the Witten field model. However,  model
has to contain an extra scalar field $Z$ providing  synchronization of
the phase transitions in $A$ and $B$ sectors.~\footnote{In fact the Morris
model contains fife complex chiral fields
$\phi _i=\{ Z, \phi _-,\phi _+, \sigma _-, \sigma _+ \} $. However,
the following identification of the fields is assumed $ \phi =\phi _+;\;
\bar \phi = \phi _-$ and $ \sigma = \sigma _+;\; \bar \sigma =\sigma_-$.
In previous notations $\phi \sim\phi_A$ and $\sigma \sim \phi _B$.}
\par
The effective Lagrangian of the Morris model has the form
\begin{eqnarray}
L=-2(D^\mu \phi )\overline {( D_\mu  \phi  )}-2(\tilde D^\mu \sigma )
(\overline {\tilde D_\mu \sigma} )-\partial ^\mu Z \partial _\mu
\bar Z \nonumber \\
-\frac 14F^{\mu \nu }F_{\mu \nu }-
\frac 14 F_B ^{\mu \nu }F_{B\mu \nu}-V (\sigma, \phi, Z),
\label{SL}
\end{eqnarray}
where the potential $V$ is determined through the superpotential $W$ as
\begin{equation}
V= \sum _{i=1} ^5 \vert W_i \vert ^2=2\vert\partial W /\partial \phi\vert^2 +
2\vert\partial W /\partial \sigma \vert^2 +
\vert\partial W /\partial Z \vert^2 .
\label{SV}
\end{equation}
The following superpotential, yielding the gauge invariance and
renormalizability of the model, was suggested
\footnote{Superpotential is holomorfic function of
$\{Z,\phi, \bar\phi, \sigma,
\bar \sigma\}$.}
\begin{equation}
W=\lambda {Z}(\sigma \bar\sigma -\eta ^2) + ( c Z+m ) \phi\bar \phi,
\label{SW}
\end{equation}
where the parameters $\lambda $, $c$, $m$, and $\eta $ are real
positive quantities.
\par
The resulting scalar potential $V$ is then given by
\begin{eqnarray}
V= & \lambda ^2(\bar \sigma \sigma -\eta ^2)^2
+2\lambda c(\bar\sigma\sigma -\eta ^2)\phi \bar\phi
+c^2(\bar\phi \phi)^2  + \\
&2 \lambda ^2 \bar Z Z \bar\sigma \sigma
+2(c\bar Z +m)(cZ+m)\bar\phi \phi.  \nonumber
\label{Vfull}
\end{eqnarray}
\subsection{Supersymmetric vacua}
>From (\ref{SV}) one sees that the supersymmetric vacuum states,
corresponding to the lowest value of the potential,
are determined by the conditions
\begin{eqnarray}
F_\sigma = - \partial \bar W / \partial \bar \sigma =0; \\
F_\phi = - \partial \bar W / \partial \bar \phi =0;\\
F_Z = - \partial \bar W / \partial \bar Z  =0,
\label{Svac}
\end{eqnarray}
and yield $V=0$.
These equatioins lead to two supersymmetric vacuum states:
\begin{equation}
I ) \qquad Z=0;\quad \phi=0 ;\quad \vert\sigma\vert=\eta ;\quad W=0;
\label{true}
\end{equation}
and
\begin{equation}
II ) \qquad Z=-m/c;\quad \sigma=0; \quad \vert \phi \vert =\eta
\sqrt{\lambda/c};\quad W=\lambda m\eta ^2/c.
\label{false}
\end{equation}
We shall take the state $I$ as a true vacuum state for external region of
the bag, and the state $II$ as a false vacuum state inside the bag.
\par
The field equations following from this Lagrangian have the
form
\begin{equation}
\tilde D_\mu \tilde D^\mu \sigma - \sigma
\left[ \lambda ^2(\bar\sigma \sigma -\eta ^2)+\lambda c\bar\phi \phi
+\lambda ^2\bar Z Z\right] =0,
\label{sig}
\end{equation}
\begin{equation}
D_\mu D^\mu \phi - \phi
\left[ \lambda c(\bar\sigma \sigma -\eta ^2)+c^2\bar\phi \phi +(c\bar Z
+m)(cZ+m)\right] =0,
\label{phi}
\end{equation}
\begin{equation}
\nabla_\mu \nabla ^\mu Z-
2\lambda ^2 Z\bar{\sigma}\sigma - 2c(cZ+m)\bar\phi \phi =0,
\label{Z}
\end{equation}
\begin{equation}
-\nabla_\nu \nabla ^\nu B_\mu =I_{B\mu }=
i2g\left[ \bar{\sigma}(\tilde D_\mu \sigma )-\sigma
(\overline{\tilde D_\mu \sigma} )\right] ,
\label{B}
\end{equation}
\begin{equation}
-\nabla_\nu \nabla ^\nu A_\mu =I_{A\mu }=
i2e\left[ \bar\phi (D_\mu \phi )-\phi (\overline {D_\mu \phi})\right],
\label{A}
\end{equation}
where $\nabla _{\mu}$ is a covariant derivative.\footnote{Factors 2 by charges
are connected with factors 2 in effective Lagrangian and can be removed by a
rescaling of the fields $\sigma$ and $\phi$.}
\par
\subsection{Gauge fields and superconducting sources}
\par
It is convenient to select amplitudes and phases of the chiral fields
\begin{equation}
\phi = \Phi e^{i\chi _A},
\sigma = \Sigma e^{i\chi _B}.
\label{PhiSig}
\end{equation}
Field $Z$ is considered as real.
By this ansatz the real part of the equation (\ref{phi})
yields
\begin{equation}
\frac{D^\mu D_\mu \Phi} \Phi -
\left[ \lambda c(\Sigma ^2 -\eta ^2)+c^2 \Phi ^2 +(cZ+m)^2+ U_A\right]=0,
\label{RePhi}
\end{equation}
where $U_A=(\chi_{A,\mu} + e A_\mu)^2 $,
and its imaginary part is
\begin{equation}
\left[\nabla ^\mu + 2(\ln \Phi),_{\mu} \right] (\chi_{A,\mu} +e A_\mu) =0.
\label{ImPhi}
\end{equation}
Similarly, one obtains the real part of (\ref{sig})
\begin{equation}
\frac{\tilde D^\mu \tilde D_\mu \Sigma} \Sigma -
\left[ \lambda c(\Sigma ^2 - \eta ^2)+c^2\Phi ^2 +(cZ+m)^2
+ U_B \right] =0,
\label{ReSig}
\end{equation}
where $U_B =(\chi_{B,\mu} + g B_\mu)^2, $
and imaginary part
\begin{equation}
\left[\nabla ^\mu + 2 (\ln\Sigma),_{\mu}\right] (\chi_{B,\mu} +e B_\mu) =0.
\label{ImSig}
\end{equation}
The equations (\ref{Z}),(\ref{B}) and (\ref{A}) take the form
\begin{equation}
\nabla_\mu \nabla ^\mu Z-
2\lambda ^2Z\Sigma ^2 - 2c(cZ+m)\Phi ^2 =0,
\label{mZ}
\end{equation}
\begin{equation}
\nabla_\nu \nabla ^\nu B_\mu =-I_{B\mu }=
4g\Sigma ^2 (\chi_{B,\mu} + g B_\mu),
\label{mB}
\end{equation}
\begin{equation}
\nabla_\nu \nabla ^\nu A_\mu =-I_{A\mu }=
4e\Phi ^2 (\chi_{A,\mu} + e A_\mu).
\label{mA}
\end{equation}
This is a very complicated system of coupled equations. However, one can
see that the behavior of the gauge fields can be studied independently in
the nearly independent sectors $A$ and $B$ when taking into account the
known vacuum states (\ref{true}) and (\ref{false}) for internal and
external regions of the bag. At the same time, as the first approximation,
the interaction of these sectors can be studied setting aside the gauge
fields. In our analysis we will follow this line.
\subsubsection{Charged superconducting sphere}
When considering sector $A$ equations
(\ref{RePhi}),(\ref{ImPhi}) can be simplified by assuming that the
amplitudes are functions of radial coordinate $r$.
Setting
$\Phi=\Phi (r)$ and $\Sigma=\Sigma (r)$,  one obtains
\begin{equation}
\Phi^{\prime \prime} +\frac 2r \Phi ^\prime -
\Phi \left[ \lambda c(\Sigma ^2 -\eta ^2)+c^2 \Phi ^2 +(cZ+m)^2+ U_A \right]
=0,
\label{rRePhi}
\end{equation}
where
\begin{equation}
U_A=(\chi_{A,\mu} + e A_\mu)^2.
\label{UA}
\end{equation}
\begin{equation}
(\partial ^\mu + 2\Phi ^\prime \vec n ^\mu) (\chi_{A,\mu} +e A_\mu) =0,
\label{rImPhi}
\end{equation}
\begin{equation}
\nabla_\nu \nabla ^\nu A_\mu =-I_{A\mu }=
4e\Phi ^2 (\chi_{A,\mu} + e A_\mu),
\label{rmA}
\end{equation}
where $\vec n^\mu = (0, x,y,z)/r$ is the unit radial vector.
Equation (\ref{rmA})
shows that field  $A_\mu$ is massless if $\Phi=0$ and acquires mass
in the region of the non-zero field $\Phi$.
As it follows from (\ref{true}),(\ref{false}) for true vacuum outside
the core we have $\Phi=0$, while inside the core
$\Phi =\eta \sqrt{\lambda/c}$, and field $A_{\mu}$ acquires
the mass $m_A= e \eta \sqrt {2 \lambda /c}$ inside the core as the
consequence of equation (\ref{rmA}). The $U(I)$
gauge symmetry is broken showing that the core is superconducting in
respect to the field $A_{\mu}$.
Indeed, it follows from
(\ref{rmA}) and (\ref{rImPhi}) that for $\Phi \ne 0 $
\begin{equation}
\partial ^\mu (\chi_{A,\mu} +e A_\mu) =
-\Phi ^\prime (\vec n ^\mu I_{A\mu})/(e \Phi ^2)=0,
\label{cc}
\end{equation}
so far as $\vec n ^\mu I_{A\mu} =I_{Ar} $ must vanish for the stationary
system as a radial component of e.m. current.
Rewriting (\ref{cc}) in the form
\begin{equation}
\partial ^\mu I_{A\mu}=0,
\label{cc1}
\end{equation}
one sees the conservation of the e.m. $A$-current corresponding to
superconductivity.
\par
For a stationary charged ball-like source field $A_{\mu}$ can be described
as $A=A_\mu dx^\mu =\frac er {\cal{A}} (r)(dt +dr)$ where the factor
$ {\cal{A}}$ describes exponential fall-off inside the superconductor.
The symmetry of the problem shows
that $I_\theta =I_r=I_\phi =0,$ and the stationary condition is $\dot I_0=0$,
that yields $(\dot \chi_{A} +e A_0)=- I_{A0}/(2e \Phi ^2) $. Therefore,
the term $U_A $ in the equation (\ref{rRePhi}) acquires the form
 $U_A=-(\omega_A + e A_0)^2, $ where $\omega_A=\dot \chi _A$.
\par
The treatment of the gauge field $B_\mu$ in $B$ sector is similar
in many respects because of the symmetry between  $A$ and $B$ sectors
allowing one to consider the state $\Sigma = \eta$ in outer region as
superconducting one in respect to the gauge field $B_\mu$.
Field $B_\mu$ acquires the mass $m_B= g \eta $ in outer region, and the
$\tilde U(I)$ gauge symmetry is broken, which provides confinement of the
$B_{\mu}$ field inside the bag. However, in the bag models
the QCD-vacuum is ``dual" to usual superconductivity
forming a magnetic superconductor \cite{SkS,dual}.
The bag can also be filled by quantum excitations of fermionic,
or non Abelian fields.
The interior space of the Kerr bag is regularized in this model since the
Kerr singularity and twofoldedness are suppressed by function $f=f_0(r)$.
However, a strong increase of the fields near the former Kerr singularity
can be retained leading to the appearance of traveling waves along the
boundary of the disc.
\par
\subsubsection{Superconducting disc-like source of
the Kerr-Newman e.m. field}
\par
A similar treament can be performed for the disc-like Kerr source if we
consider the $r$-coordinate as ellipsoidal coordinate of the Kerr
geometry (\ref{ell}). The exact treatment on the Kerr-Newman background is
rather complicated and will be given elsewhere \cite{bur3}. Because of that,
we consider here the case corresponding to small values of the mass
and charge parameters taking into account the $e/a$ and $m/a$ ratios
(\ref{par}).
The Kerr coordinates $r,\theta, \phi$ are connected with
Cartesian coordinates $(x,y,z)$ by relations
\begin{eqnarray}
x+iy =& (r+ia) e^{i\phi} \sin \theta, \\
z =& r\cos \theta.
\label{coord}
\end{eqnarray}
In the considered limit we have in the Kerr coordinates
\begin{eqnarray}
\nabla ^\mu \nabla _\mu =
\frac 1{r^2 + a^2 \cos ^2 \theta} \lbrack \partial _r (r^2 + a^2) \partial _r
+ \frac 1{\sin \theta}\partial _\theta \sin \theta \partial _\theta \\
+ \frac 1{\sin ^2 \theta}\partial ^2 _\phi - \partial ^2_t +
 2a \partial _r \partial _\phi \rbrack.
\label{Del}
\end{eqnarray}
The surfaces $r=const.$ represent oblated spheroids and $\theta = const.$ are
confocal hyperboloids.
By assuming that the scalar amplitudes are only functions of $r$,
it is reduced to
\begin{equation}
\nabla ^\mu \nabla _\mu =
\frac1{r^2 + a^2 \cos ^2 \theta} \partial _r (r^2 + a^2) \partial _r .
\label{RDel}
\end{equation}
In this case the
equations (\ref{RePhi}),(\ref{ReSig}) and (\ref{mZ}) take the form
\begin{equation}
\frac 1{r^2 + a^2 \cos ^2 \theta}(\Phi^{\prime \prime} + 2r \Phi ^\prime) -
\Phi \left[ \lambda c(\Sigma ^2 -\eta ^2)+c^2 \Phi ^2 +(cZ+m)^2+ U_A \right]
=0,
\label{kRePhi}
\end{equation}
where $U_A=(\chi_{A,\mu} + e A_\mu)^2; $
\begin{equation}
\frac 1{r^2 + a^2 \cos ^2 \theta}(\Sigma^{\prime \prime} +
2r \Sigma^\prime) - \Sigma \left[ \lambda c(\Sigma ^2 -
\eta ^2)+c^2\Phi ^2 +(cZ+m)^2 \right] =0,
\label{rReSig}
\end{equation}
where $U_B=g^{\mu\nu}(\chi_{B,\mu} + g B_\mu)(\chi_{B,\nu} + g B_\nu); $
\begin{equation}
\frac 1{r^2 + a^2 \cos ^2 \theta}(Z^{\prime \prime} + 2r Z^\prime) -
\lambda ^2Z\Sigma ^2 - c(cZ+m)\Phi ^2 =0,
\label{rmZ}
\end{equation}
and by $r\ll a\cos \theta $ the equations
tend to ones with a planar symmetry depending only on coordinate
$z=r\cos \theta$. Such a region takes the place near the axis of the
disc-like source where the solution tends to a planar domain wall.
Another limiting case corresponds to the region near the Kerr singularity
when $r\ll a$ and $\cos \theta = 0$. Delambertian in this region takes
the form
\begin{equation}
\frac 1{r^2}(a\partial ^2_r + 2 \partial_r).
\label{sing}
\end{equation}
Finally, in the region  $r\gg a $ Delambertian takes the usual form
corresponding to the spherical symmetry of the problem.
Let us now consider specifical behavior of the Kerr-Newman e.m. field
surrounding the disc-like superconducting core.
\par
In our approximation contravariant form of metric in Kerr coordinates is
\begin{equation}
g^{\mu\nu} =
\left( \begin{array}{cccc}
-1&0&0&0 \\
0&\frac{r^2 +a^2}{\Sigma}&0&\frac a{\Sigma} \\
0&0&\frac 1{\Sigma}&0 \\
0&\frac a{\Sigma}&0&\frac 1{\Sigma \sin ^2 \theta}
\end{array} \right),
\label{gcont}
\end{equation}
and
\begin{equation}
\sqrt {-g} = \Sigma \sin \theta.
\label{det}
\end{equation}
>From physical reason and from the symmetry of the problem
we set the constraints that current must vanish in radial
$\vec n=(0,1,0,0)$ and angular $\vec e_\theta =(0,0,1,0)$
directions
\begin{equation}
I^r =(I\vec n)=0; \qquad I^\theta =(I\vec e_\theta)=0.
\label{Inth}
\end{equation}
By using (\ref{gcont}) one obtains
\begin{equation}
\chi_{,\theta} = -e A_\theta,
\label{Itheta}
\end{equation}
and
\begin{equation}
I_r =  = - \frac a {r^2 + a^2} I_\phi.
\label{Ir}
\end{equation}
Taking into account (\ref{det}), (\ref{gcont}) and (\ref{Inth})
and stationarity $\partial _0 I^\mu =0$,
the equation of current conservation (\ref{cc}) is reduced to the form
\begin{equation}
D_\mu I^\mu =\partial _\phi I^\phi =0,
\label{Iphi}
\end{equation}
which yields
\begin{equation}
\chi _{,\phi \phi} = 0.
\label{chi2phi}
\end{equation}
Vector potential of the Kerr-Newman e.m. field has the form
\begin{equation}
A_{KN}= \frac {er}\Sigma (dt +dr - a \sin ^2 \theta d\phi).
\label{AKN}
\end{equation}
For the Kerr-Newman field connected with superconducting source we retain
this form adding an extra factor  ${\cal A}(r)$ that describes fall-off of
the field inside the source
\begin{equation}
A= {\cal A}(r)\frac {er}\Sigma (dt +dr - a \sin ^2 \theta d\phi).
\label{AsKN}
\end{equation}
Since $A_\theta =0$ we have from (\ref{Itheta})
\begin{equation}
\chi_{,\theta}=0,
\label{chitheta}
\end{equation}
and consequently
\begin{equation}
\chi_{,r\theta}=\chi_{,\phi \theta}= 0 .
\label{chirthe}
\end{equation}
Together with (\ref{chi2phi}) and condition of stationarity $\partial _0 I^0$
it  gives the integral
\begin{equation}
\chi=n\phi + \omega t + \chi_{1}(r),
\label{tnchi}
\end{equation}
where $n$ and $\omega$ are constants, and $n$ has to be integer to provide
single-valuedness of phase $\chi$.
Since $\chi_{,r}=\chi _1^\prime (r)$ one obtains from (\ref{Itheta})
\begin{equation}
\chi_{1}^\prime= - \frac{an + e{\cal A}(r)}{r^2 +a^2},
\label{dchi1}
\end{equation}
that can be integrated leading to the expression
\begin{equation}
\chi (r,t,\phi)=n\phi +\omega t - \int \frac{an +e{\cal A}(r)}{r^2+a^2} dr.
\label{finchi}
\end{equation}
Now one can calculate the extra contribution to scalar potential $U_A$
caused by e.m. field (\ref{UA})
\begin{equation}
U_A= \frac 1{(r^2 +a^2) } ( n/\sin ^2 \theta - a {\cal A}(r)/\Sigma )^2 -
(\omega + e {\cal A}(r)/\Sigma )^2.
\label{KerrUA}
\end{equation}
One sees that by $n\ne 0$ potential depends from $\theta$,  and moreover, it
is singular at $\theta =0$ that corresponds to the appearance of the Dirac
monopole string.
\par
\subsection{Supersymmetric domain wall and phase transition}
\par
Now we would like to analyze a supersymmetric domain wall based on the
Morris field model. It is described by the system of equations
(\ref{RePhi}), (\ref{ReSig}), (\ref{mZ}).
\par
To simplify the problem we consider the planar approximation and neglect
the contributions to potential $U_A$ and $U_B$ resulting from the gauge
fields.
The supersymmetric Lagrangian for the chiral matter fields ( bosonic sector )
has the form
\begin{equation}
(-g)^{-1/2}L_{matB} =- K_{i\bar j}(D_{\mu} \Phi ^i)\overline
{ (D^{\mu}\Phi ^j)} - V;
\label{Lmat}
\end{equation}
where
\begin{equation}
V=K^{i\bar j}(D_{i} W)\overline {(D_{j} W )},
\label{pot}
\end{equation}
$K_{i\bar j}$ is K\"ahler metric,
$D_i= \partial/\partial \Phi  + \bar \Phi ^i$, and
$W$ is a superpotential.
Stress-energy tensor is given by
\begin{equation}
T_{\mu \nu} =-\frac{1}{\sqrt{-g}}\frac{\partial
L_{mat}}{\partial g^{\mu \nu}} =
K_{i\bar j}(D_{\mu} \Phi ^i)\overline {(D_{\nu}\Phi ^j)}
-\frac 12 g_{\mu \nu}
[ K_{i\bar j}(D_{\lambda} \Phi ^i)\overline {(D^{\lambda}\Phi ^j)} + V].
\label{T}
\end{equation}
In our case the set of fields $\Phi ^i =
(Z,\Sigma_+,\Sigma _-, \Phi _+, \Phi _-)$,
the superpotential  $W$ is
\begin{equation}
W = \lambda Z(\Sigma_+ \Sigma _- - \eta ^2) +(cZ+m)\Phi _+ \Phi _-,
\label{W}
\end{equation}
 and $K_{i,\bar j}=\delta_{i \bar j}$.
\par
For the planar domain wall the fields depend on the coordinate $z$ only,
and, neglecting the contributions from gauge fields, one can consider the
fields as real.
The non-zero components of the stress-energy tensor take the form
\begin{eqnarray}
T_{00} & = &-T_{xx} =- T_{yy}=\frac{1}{2}[ \delta _{ij}( \Phi ^i,_z)
(\Phi ^j,_z) + V];\\
T_{zz}& = &\frac{1}{2}[ \delta_{ij}( \Phi ^i,_z)(\Phi ^j,_z) - V].
\label{Tflat}
\end{eqnarray}
Since in our case the potential is positive $V\ge 0$, the problem permits a
mathematical analogy
with the problem of motion of a particle of unit mass propagating in the
potential
$U_{part}=-V(\Phi ^i)$ in a space-time having
the  space coordinates $\Phi ^i$ and time coordinate $z$, see \cite{Raj}.
The field equations, following from (\ref{Lmat}) in this case
\begin{equation}
2 \Phi ^i ,_{zz} - V,_{\Phi _i} =0,
\label{feq}
\end{equation}
lead to the equation
\begin{equation}
2 \Phi _i ,_z \Phi ^i ,_{zz} - V,_{\Phi _i} \Phi _i ,_z=0,
\label{sum}
\end{equation}
that can be represented in the form
\begin{equation}
 \partial _z[(\Phi ^i ,_z)^2  - V] =0.
\label{res}
\end{equation}
Integral of this equation is
\begin{equation}
 \frac{1}{2}(\Phi ^i ,_z)^2  + U_{part} =const. ,
\label{part}
\end{equation}
where $U_{part} =-\frac 12 V$ an effective potential for the dynamics of the
fictitious particle, and (\ref{part}) describes conservation of its
kinetic and potential energy.
Since the wall interpolates between two disconnected vacuum regions,
we have asymptotic behavior
\begin{equation}
U_{part}\vert _{z=\pm \infty}=
-\frac 12 (\Phi ^i ,_z)^2 \vert _{z=\pm \infty}=0,
\label{inf}
\end{equation}
that yields $const.=0$ and
\begin{equation}
V = (\Phi ^i ,_z)^2.
\label{U}
\end{equation}
As a result of this equation we obtain
\begin{eqnarray}
T_{zz}& = &0;\\
T_{00} & = &-T_{xx} =- T_{yy}= ( \Phi ^i,_z)^2 = V.
\label{Tres}
\end{eqnarray}
Therefore, similarly to the simplest kink solution for domain wall
(\ref{kink}),(\ref{Tkink}) the wall
has a positive tangential stress which is equal to energy density and zero
pressure in the transverse to wall direction. It should be noted that this
property of the wall has been derived without obtaining an explicit
solution of the field equations and without specifying
the form of potential $V$ on the basis of only its positivity and boundary
conditions (\ref{inf}).
\par
Let us now assume that the wall has a spherical form with a small
thickness in respect to the radius.  One can calculate a
gravitational mass produced by the spherical wall in outer region.
Using the
Tolman relation $M =\int dx^3 \sqrt{-g}(-T_0^0+T_1^1 +T_2^2 +T_3^3)$,
replacing coordinate $z$ on radial coordinate $r$, and integrating over
sphere one obtains
\begin{equation}
M_{bubble}= -4\pi \int V(r) r^2 dr = -4\pi \int (\Phi ^i,_r)^2 r^2 dr .
\label{Mwall}
\end{equation}
The resulting effective mass is negative, which is caused by gravitational
contribution of the tangential stress. The repulsive
gravitational field was obtained in many singular and
smooth models of domain walls \cite{IpsSik,lop2,CvSol,CG}.
In particle models based on singular charged bubbles, this
negative gravitational energy of domain wall cancels the gravitational
part of the external electromagnetic field energy \cite{lop2}.
One should note, that similar gravitational contribution to the mass caused
by interior of the bag will be $M_{gr.int}=\int D  r^2 dr
=-\frac 23 \Lambda r_0^3$, and it depends on the sign of curvature inside
the bag.
\par
An additional information can be extracted by the choice of a reasonable
approximation to optimal trajectory of the phase transition from vacuum
state I ( $z=-\infty$ ) to vacuum state II ( $z=+\infty$ ) in the space of
fields $\Phi ^i (z) = \lbrace Z(z), \Phi (z) , \Sigma (z) \rbrace $ with a
possible minimization of the potential \cite{Raj}.
The potential (\ref{Vfull}) for real fields takes the form
\begin{equation}
V= \lambda ^2 \lbrack (\Sigma ^2+ \Phi ^2 -\eta ^2 )^2 +
 Z^2 \Sigma ^2 +( Z +m/\lambda)^2 \Phi ^2 \rbrack ,
\label{Vreal}
\end{equation}
where for simplicity we set $c=\lambda$ and the position of domain wall
$z_0=0$.
To cancel the first term of this expression one can set the following
parametrization for $\Phi$ and $\Sigma$:
\begin{equation}
\Sigma = \eta \sin \beta,
\Phi = \eta\cos \beta.
\label{tr}
\end{equation}
The subsequent minimization of the next two terms yields the
parametrization  for $Z$
\begin{equation}
Z(z)=-(m/\lambda) \cos^2 \beta .
\label{Zz}
\end{equation}
The phase transition occurs for $\beta \subset [0,\pi/2]$.
The resulting dependence of the potential on $z$ will be
\begin{equation}
V_{opt}(\beta)= \eta ^2 m^2\sin ^2 2\beta /4.
\label{V0}
\end{equation}
Substituting (\ref{tr}),(\ref{Zz}) and (\ref{V0}) in (\ref{U}) one obtains
for dependence $\beta (z)$ the equation
\begin{equation}
d\beta / dz=  { m\sin 2\beta}/\sqrt {1 +
(m^2/\lambda ^2 \eta ^2)\sin ^2 2\beta}.\label{beta}
\end{equation}
\par
\subsubsection{Bogomol'nyi transformation, BPS bound and total mass of
the bag}
\par
Starting from the expressions (\ref{Tflat}) and (\ref{pot}) by
$K_{i,\bar j}=\delta_{i \bar j}$ one can use the Bogomol'nyi
transformation \cite{Bog} and represent the energy density as follows
\begin{eqnarray}
\rho &=& T_{00} =\frac{1}{2} \delta _{ij}\lbrack ( \Phi ^i,_z)
(\Phi ^j,_z) + (\frac {\partial W}{\partial \Phi ^i})
(\frac {\partial W}{\partial \Phi ^j})\rbrack \\
 &=&\frac{1}{2} \delta _{ij}\lbrack  \Phi ^i,_z + \frac {\partial W}{\partial
\Phi ^j}\rbrack
\lbrack  \Phi ^j,_z + \frac {\partial W}{\partial\Phi ^i}\rbrack
- \frac{\partial W}{\partial \Phi ^i}\Phi ^i,_z,
\label{trbog}\end{eqnarray}
where the last term is full derivative.
Then, integrating over the wall depth $z$ one obtains for the surface
density of the wall energy
\begin{equation}
\epsilon=\int _0 ^\infty \rho dz =\frac 12 \int\Sigma _i
( \Phi ^i,_z + \frac {\partial W}{\partial\Phi ^i})^2 dz + W(0) -W(\infty).
\label{edens}
\end{equation}
The minimum of energy is achieved by solving the first-order Bogomol'nyi
equations $\Phi ^i,_z + \frac {\partial W}{\partial\Phi ^i} =0$, or
in terms of $Z, \Phi, \Sigma $
\begin{eqnarray}
 Z^\prime &=& -  \lambda(\Sigma ^2-\eta ^2) - c \Phi ^2, \\
 \Sigma ^\prime &=& -\lambda Z\Sigma, \\
\Phi ^\prime &=& - (cZ +m) \Phi.
\label{bogeq}
\end{eqnarray}
Its value is given by
$\epsilon = \epsilon _{min} =W(0)-W(\infty)=\lambda m \eta^2/c$.
Therefore, this domain wall is BPS-saturated solution.
One can see that energy is proportional to parameter $m$ which determines
amplitude of variation of the field $Z$. Therefore,
the additional field $Z$, which appears only in the supersymmetric version
of the model, plays an essential role for the formation of this domain wall.
Substituting approximate solution
(\ref{tr}),(\ref{Zz}),(\ref{beta}) in Bogomol'nyi equations one sees that
this approximation neglects the term $(Z^\prime)^2 $.
\par
By using the relations (\ref{Tres}),(\ref{Mwall}) and (\ref{edens})
one obtains that the total energy of a uncharged bubble
forming from the supersymmetric BPS saturated domain wall is
\begin{equation}
E_{0 bubble} = E_{wall} =
4\pi \int _0 ^\infty \rho r^2 dr \approx 4\pi r_0 ^2
\epsilon_{min},
\label{E0tot}
\end{equation}
where $r_0$ is radius of the bubble.
Corresponding total mass following from the Tolman relation will be
negative
\begin{equation}
M_{0 bubble} = - E_{wall} \approx -4\pi r_0 ^2 \epsilon_{min}.
\label{M0tot}
\end{equation}
It is the known fact showing that the uncharged bubbles are unstable and
form  the time-dependent states \cite{CvSol,CG}.
\par
For charged bubbles there are extra positive terms:
contribution caused by the energy and mass of the external electromagnetic
field
\footnote{ This analysis represents only a qualitative estimation
in the spirit of the typical treatments of the bag models.
There can also be other contributions to the total mass and
energy, in particular, the energy of the interior of the bag and
the energy of the tail of electromagnetic field
penetrating in superconducting core, which can be comparable with
$E_{e.m.}$.}
\begin{equation}
E_{e.m.} = M_{e.m.} = \frac{e^2}{2r_0},
\label{EMem}
\end{equation}
and contribution to mass caused by gravitational field of the external
electromagnetic field ( determined by Tolman relation for
the external e.m. field)
\begin{equation}
M_{grav. e.m.} =  E_{e.m.} = \frac{e^2}{2r_0}.
\label{Mgrem}
\end{equation}
As a result the total energy for charged bubble is
\begin{equation}
E_{tot.bubble} = E_{wall} + E_{e.m.} = 4\pi r_0 ^2 \epsilon_{min}
+ \frac{e^2}{2r_0},
\label{Etot}
\end{equation}
and the total mass will be
\begin{equation}
M_{tot.bubble} = M_{0 bubble} + M_{e.m.} + M_{grav.e.m.} =
- E_{wall} + 2E_{e.m.} = -4\pi r_0 ^2 \epsilon_{min} + \frac{e^2}{r_0}.
\label{Mtot}
\end{equation}
Minimum of the total energy is achieved by
\begin{equation}
r_0=(\frac{e^2}{16\pi \epsilon_{min}})^{1/3},
\label{rmin}
\end{equation}
which yields the following expressions for total mass and energy of
the stationary state
\begin{equation}
M_{tot}^* =E_{tot}^* =\frac {3e^2}{4r_0}.
\label{EMst}
\end{equation}
One sees that the resulting total mass of charged bubble is positive,
however, due to negative contribution of $M_{0 bubble}$ it can be lower
than BPS energy bound of the domain wall forming this bubble.
This remarkable property of the bubble models ( called as `ultra-extreme'
states for the Type I domain walls in \cite {CvSol} ) allows one to overcome
BPS bound \cite{GH} and opens the way to get the ratio $m^2 \ll e^2$ which is
necessary for particle-like models.
\par
\subsubsection{Supergravity domain wall}
\par
In supergravity the scalar potential has a more complicate form
\cite{WB,CvSol,CG,Mor2}
\begin{equation}
V_{sg}=e^{k^2 K}( K^{i\bar j}D_{i} W\overline {D_{j} W }
-3k^2 W \bar W),
\label{SGpot}
\end{equation}
where $K$ is K\"ahler potential $ K^{i\bar j}=\frac {\partial ^2 K}
{\partial \Phi _i \partial \bar \Phi _j}$, and
$k^2=8\pi G_N$, $G_N$ is the Newton constant.
For small values of $kW$, this expression
turns into potential of global susy (\ref{pot}) and the above treatment
of the charged domain wall bubble is valid in supergravity in this
approximation.
However, in the first order in $k^2$ the internal vacuum state II)
does not preserve supersymmetry since  $W= \lambda m \eta^2/c$ inside
the bag, and $D_i W \approx k^2 K_i W \ne 0$ there.
Besides, there apprears an extra contribution to stress-energy tensor
having the leading term
\begin{equation}
T_{\mu \nu} =3(k^2/8\pi) e^{k^2K}\vert W \vert ^2 g_{\mu\nu},
\label{Tsg}
\end{equation}
and yielding the negative cosmological constant $\Lambda =-3 k^4 e^{k^2K}
\vert W \vert ^2$ and to anti-de Sitter space-time for the bag interior.
General expression for cosmological constant inside the bag has the form
\begin{equation}
\Lambda=k^4 e^{k^2 K}\sum _{i}\{k^2 | K_i W |^2 - 3 |W|^2 \},
\label{lam}
\end{equation}
and leads to AdS vacuum if $k^2 | K_i W |^2 - 3 |W|^2 <0 $.
\par
In the same time the vacuum state I) in external region has $W=0$, it
preserves supersymmetry for strong values of the chiral fields and does not
give  contribution to stress-energy tensor.

\par
\subsection{Conclusion}
\par
 The analysis of the stress-energy tensor of the Kerr source in the
Kerr-Schild
class of metrics shows that the simple and natural demands of the regularity
of the energy density of the source lead to the conclusion concerning a
bag-like structure of the core of the Kerr-Newman solution. The boundary of
the bag
represents a smooth ( thick ) domain wall with a tangential stress playing
the role of a Poincar\`e stress of the core. For the thin domain wall this
domain wall can be considered as being in a rigid relativistic rotation.
\par
Among the known field models suggested for the description of extended
objects such as strings, domain walls and bubbles, only the Witten
$U(I)\times\tilde U(I)$ field model based on two gauge and two Higgs fields
can provide the long range electromagnetic external field which is necessary
for modelling the Kerr bag-like superconducting source.
\par
 The application of Morris supersymmetric version of this model to
the bag-like source
was considered, and it was shown that this model provides a phase transition
between two disconnected supersymmetric vacua. The model possesses two
superconducting sectors and two vacuum states: the vacuum state inside the
bag is superconducting in respect to the external Kerr-Newman electromagnetic
field, while the vacuum of the external region is a standard QCD-vacuum of
the bag models which can be considered as a magnetic superconductor in
respect to the inhabiting the bag quantum gauge fields.
These superconducting sectors
are almost independent and interact only via a special scalar field $Z$
providing a synchronization of phase transition.
The Bogomol'nyi equations are obtained showing that the domain wall is
BPS saturated solution.
Due to negative gravitational contribution of the domain wall to the
mass of the charged bag the total mass of the bag can be lower than
BPS energy bound of this domain wall, which represents a remarkable
property of the bubble-like core opening the way for particle-like
models with a close to real parameter ratio.
\par
The case N=1 supergravity is considered, and it is shown that strong chiral
fields can lead to a negative cosmological constant and AdS-geometry inside
the bag.
\par
The current  analysis can be considered only as a preliminary one. The
complexity of
the model does not allow one to hope to obtain an analytical solution of
the full system of equations for this model, and numerical calculations must
be used. However, one can expect essential
further progress in obtaining analytical solutions for some sectors of this
model. In particular, exact solutions for the Kerr-Newman electromagnetic
field generated by the
superconducting disc-like core on the Kerr-Newman background
apparently can be obtained, as well as the existence of the exact analytical
solutions for quantum excitations of the bosonic and fermionic fields inside
the Kerr disc-like bag on the regularized Kerr-Schild background can be
assumed.
\par
\section{Acknowledgement}
\par
We would like to thank H. Nicolai, M. Vasiliev, Yu. Obukhov, G. Magli,
G. Alekseev, S. Duplij,
E. Elizalde, S.R. Hildebrandt, J. Morris and D. Bazeia for very useful
discussions and comments.
\par

\end{document}